# High Quality Factor Silicon Cantilever Driven by PZT Actuator for Resonant Based Mass Detection

J.Lu[1], T.Ikehara[1], Y.Zhang[1], T.Mihara[2], T.Itoh[1], R.Maeda[1]

[1] National Institute of Advanced Industrial Science and Technology (AIST), 1-2-1 Namiki, Tsukuba, Ibaraki, 305-8564, Japan
[2] Future Creation Lab., Olympus Corporation, 2-3-1 Nishi-Shinjuku, Shinjuku-ku, Tokyo, 163-0914, Japan

*Abstract-* **A high quality factor ($Q$-factor) piezoelectric lead zirconat titanate (PZT) film actuated single crystal silicon cantilever was proposed in this paper for resonant based ultra sensitive mass detection. Intrinsic energy dissipation and other negative effects from PZT film were successfully compressed by separating the PZT actuator from the resonant structure. Excellent $Q$-factor, which is comparable to silicon only cantilever and several times larger than latest reported $Q$-factor of integrated cantilevers, was successfully obtained under both atmospheric pressure and reduced pressures. For a 30 μm-wide 100 μm-long cantilever, $Q$-factor was measured as high as 1113 and 7279 under the pressure of 101.2 KPa and 35 Pa, respectively. Moreover, it was found that high mode vibration can be achieved for the pursuit of great $Q$-factor. However, support loss became significant because of the increase of PZT actuator's vibration amplitude. Therefore, an optimized structure using node-point actuation was discussed and suggested herein.**

## I. INTRODUCTION

Resonant based microcantilever has been widely studied in bio-science and nano technology as an alternative to quartz crystal microbalance (QCM) sensor for molecular recognition [1, 2], virus detection [3], and single cells or nano particles weighing [4, 5]. Due to remarkable increased cantilever surface area to volume ratio by several orders, ultra-high mass detection sensitivity of up to femto-gram ($10^{-15}$ g) has been achieved in recent publications [6]. Sensitivity of the microcantilever mass sensor can be further improved by using networked sensor system or parallel sensor technology to extract specific information from every node [7]. Therefore, to be integrated with complementary metal oxide semiconductor (CMOS) circuits as well as other micro electromechanical systems (MEMS) components, piezoelectric lead zirconate titanate (PZT) cantilever is more attractive than other types, such as electrostatic [6] and electromagnetic [8] resonator, for its self-actuation self-sensing capability and unique characteristics of ultra-low actuation voltage and low power consumption.

Simultaneous actuation and sensing requires PZT-electrode stack to be integrated on cantilever. Our recent efforts reveal that energy dissipation from intrinsic loss of the PZT film was comparable to air dumping under atmospheric pressure. Under reduced pressures, energy dissipation from PZT film became predominant [9, 10]. It greatly degraded sensitivity of the cantilever via quality factor ($Q$-factor), since $Q$-factor determines minimum detectable frequency shift $\Delta f$ after loading-mass $\Delta m$ adsorption as indicated by equation (1):

$$\Delta m / m_0 = -2\Delta f / f_0 \propto 1/Q \quad (1)$$

where $m_0$ is the mass of the resonator and $f_0$ is the resonant frequency. Besides, weak output from PZT film requires high precision pre-amplifier for the detection of the resonant frequency. It leads to high cost and additional integration difficulties. Residual stress in PZT-electrode stacks is another intractable problem for system integration. It places an initial strain in PZT film, results in large cantilever curvature, and induces unexpected vibration mode.

We have already demonstrated that $Q$-factor and sensitivity of the cantilever can be improved by partially covering the cantilever with PZT film [11]. In this paper, a PZT actuated single crystal silicon cantilever, which has excellent $Q$-factor, was sucessfully developed by separating the PZT actuator from the resonant structure to suppress energy dissipation and other negative effects. Resonant frequency of the cantilever can be detected by an on-chip integrated piezoresistive gauge for better compatibility with CMOS circuits. $Q$-factor was studied under both atmospheric pressure and reduced pressures, and the energy dissipation mechanism was analyzed by comparing the measured $Q$-factor with theoretical calculation. High mode vibration was also investigated and an optimized structure using node-point actuation was discussed and suggested thereafter.

## II. DESIGN AND EXPERIMENTAL

### A. Design and Simulation

Figure 1 shows schematic view of the proposed structure.

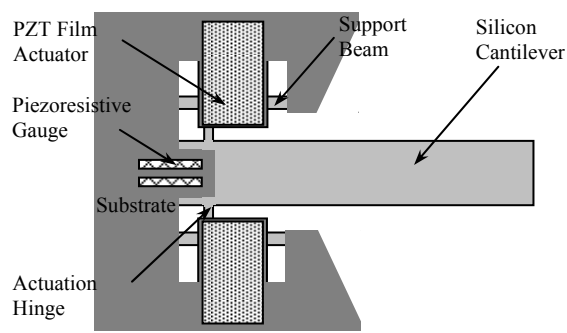

Fig. 1 Schematic view of the integrated microcantilever mass sensor.

PZT actuators (50×65 μm) were designed on both side of the silicon cantilever and connected to the cantilever via thin hinges (4×10 μm). Wheatstone bridge piezoelectric gauge was integrated close to fixed end of the cantilever to detect the resonant frequency. According to finite-element analysis (FEA) using ANSYS® simulation as shown in Fig. 2, the deformation of the actuator after apply electric voltage to PZT film can be effectively forwarded to the cantilever via actuation hinges. ANSYS® harmonic analysis reveals that after using support beam (10×15 μm) on each side of the PZT actuator, vibration amplitude of the actuator can be restricted to only a few nm to compresses its energy dissipation. Another function of the support beam is to reduce the cantilever's initial curvature.

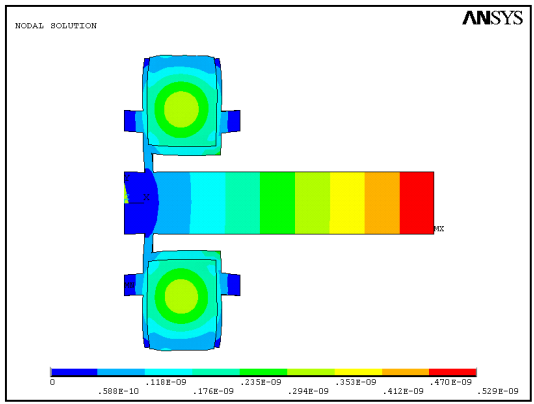

Fig. 2 Finite-element simulated cantilever deformation after apply electric voltage to PZT film.

$Q$-factor of a resonator is inversely proportional to the rate of energy dissipation. It can be estimated theoretically using equation (2):

$$\frac{1}{Q} = \frac{1}{Q_{air}} + \frac{1}{Q_{Sup}} + \frac{1}{Q_{TED}} + \frac{1}{Q_{Others}} \quad (2)$$

where $Q_{air}$ is the air damping $Q$-factor, $Q_{sup}$ is the support loss $Q$-factor, and $Q_{TED}$ is the thermoelastic loss (TED) $Q$-factor. $Q$-factor of other energy losses (e.g. PZT loss, surface loss) is denoted by $Q_{others}$.

According to analytical fluidic solution for a vibrating sphere, $Q_{air}$ of a vibrating cantilever with length $L$, width $W$ and thickness $t$ can be calculated as follows [12, 13]:

$$Q_{air} = \frac{k_n^2}{12\pi\sqrt{3}} \cdot \sqrt{\rho \cdot E} \cdot \frac{W \cdot t^2}{L \cdot R \cdot (1 + R/\delta) \cdot \mu} \quad (3)$$

in which $k_n$ is the vibration mode coefficient, $\rho$ is the density of the cantilever, $E$ is the Young's modulus of the vibrating material, $\mu$ is the viscosity of the surrounding air, and $R$ is the effective sphere radius.

$Q_{sup}$ of a cantilever can be estimated using 2-dimensional elastic wave theory with the assumption that cantilever thickness $t$ is much smaller than wavelength of the elastic wave propagation [14]:

$$Q_{sup} = C \cdot (L/t)^3 \quad (4)$$

$C$ ranges from 2.081 to near 0, depending on resonant mode.

TED arises from the coupling of mechanical stain to heat flow in a vibrating material. $Q_{TED}$ can be expressed approximately by equation (5) as follows [15, 16]:

$$Q_{TED} = \frac{C_V}{E \cdot a^2 \cdot T_0} \cdot \frac{1 + (\omega \cdot \tau)^2}{\omega \cdot \tau} \quad (5)$$

in which $\alpha$ is the thermal expansion coefficient, $T_0$ is the nominal equilibrium temperature, $C_V$ is the heat capacity at constant volume, $\omega$ is the angular resonant frequency, and $\tau$ is the relaxation time.

Air damping, support loss and TED are the principle loss mechanisms in a single layered silicon cantilever [17]. In this paper, geometries of the resonant cantilever were optimized using above theories for the pursuit of great mass detection sensitivity. The thickness $t$ of the cantilever was set at 5 μm. The width $W$ of the cantilever was set at 30, 50 and 90 μm. And the length $L$ of the cantilever was varied at 100, 150, 200, 300, and 500 μm to investigate its energy dissipation mechanism.

*B. Fabrication Process*

The device fabrication began with a silicon on insulator

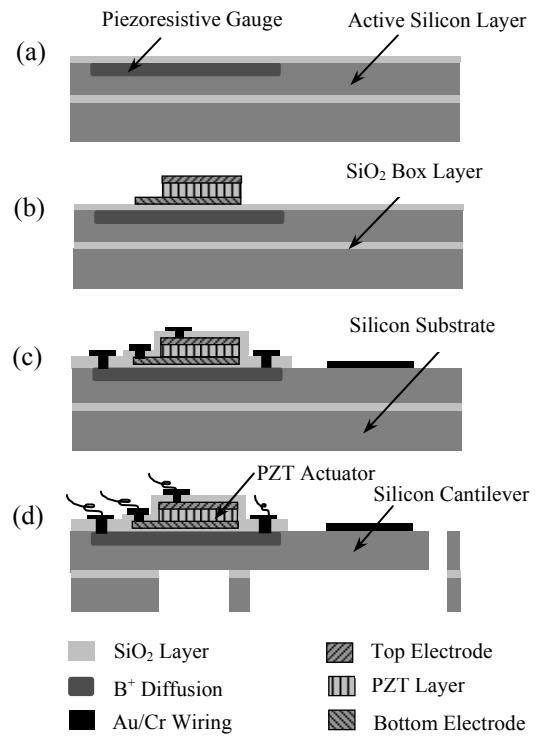

Fig. 3 Flow chart of the cantilever fabrication process.

(SOI) wafer. Cantilever thickness was determined by active silicon thickness of the SOI wafer, which was chosen as 5 μm. Fig. 3 shows flow chart of the fabrication process.

First, a 300 nm-thick $SiO_2$ layer was grown on the wafer by thermal oxidation as a passivation layer for boron diffusion. Boron was diffused at 950 °C for 160 min to make piezoresistive gauge. Then another 300 nm-thick $SiO_2$ passivation layer was grown by thermal oxidation (as shown in Fig. 3 (a)).

Second, a 250/50 nm-thick Pt/Ti bottom electrode, a 1 μm-thick PZT film, and a 30/150/30 nm-thick Cr/Au/Cr top electrode were deposited using sputter and sol-gel techniques as described in our previous report [18]. Then the sandwiched structure was etched using Argon plasma to from PZT-electrode stack (as shown in Fig. 3 (b)).

After that, a 300 nm-thick $SiO_2$ layer was deposited by sputter to passivate the PZT-electrode stack. Reactive ion etching (RIE) was used to produce electric contact holes as well as to remove the $SiO_2$ layer from the resonant structure. Then a 240/30 nm-thick Au/Cr layer was deposited by sputter and etched by Argon plasma to create electric wiring. At the same time, an Au/Cr pattern was formed on the cantilever as an area for polymer or bio-chemical adsorbent deposition (as shown in Fig. 3 (c)).

Finally, the cantilever and the actuator was defined using inductively coupled plasma (ICP) RIE from top side of the wafer, and then released from back side of the wafer using ICP-RIE and buffered hydrofluoric acid (BHF) solution to remove silicon substrate and $SiO_2$ box layer, respectively (as shown in Fig. 3 (d)).

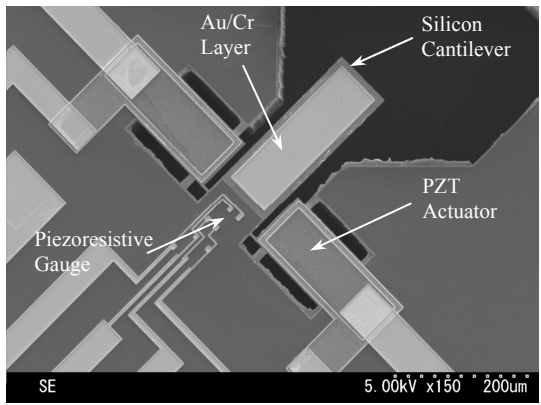

Fig. 4 SEM image of a fabricated 90 μm-wide 300 μm-long cantilever.

Figure 4 shows scanning electron microscope (SEM) image of a fabricated microcantilever. By taking advantageous of the support beam and the stress-free active silicon layer in SOI wafer, the whole structure was flat and the cantilever exhibits negligible initial curvature as preferred. However, in a conventional multi-layered PZT cantilever with the same length, tip displacement can be as high as 80~100 μm as we demonstrated before [19].

### C. Q-factor Evaluation

To compare the proposed structure with other cantilevers, $Q$-factors of the cantilever were measured using a laser Doppler vibrometer (LDV, LT7901, Graphtec Co.) and a network analyzer (HP4395A, Agilent Technologies Inc.) as shown in Fig. 5. Details have been reported elsewhere [10]. The measurement system was set on a floating table to isolate mechanical noise.

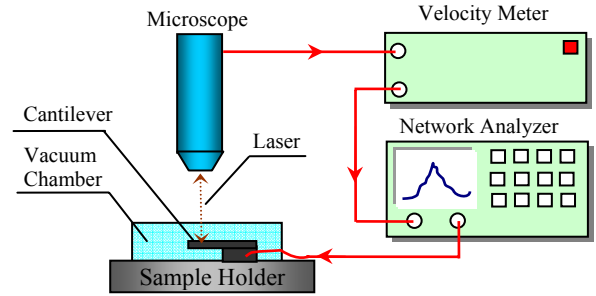

Fig. 5 Schematic view of the cantilever $Q$-factor measurement system.

### III. RESULTS AND DISCUSSION

#### A. Under Atmospheric Pressure

Solid marks in Fig. 6 shows measured $Q$-factors of the cantilever at different lengths and widths under atmospheric pressure. The calculated $Q$-factor of Au/Cr covered silicon cantilever after considering air damping, support loss and TED effect using equation (2) were plotted in Fig. 6 too by hollow marks for comparison. The material properties used in calculation can be found in reference [20]. Clearly, all the cantilevers exhibit excellent $Q$-factor, which is several times larger than that of conventional PZT cantilever and other latest reported integrated cantilevers [8, 10, 21, 22]. The measured *Q*-factor fitted well with theoretical calculation, indicating that energy dissipation from the PZT film and the actuator was dramatically suppressed after separating the PZT actuator from the resonant structure. *Q*-factor of the cantilever is therefore comparable with single layered silicon cantilevers as exhibited in reference [13].

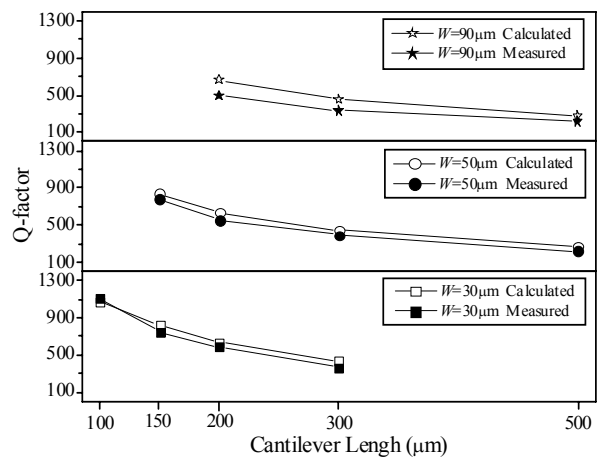

Fig. 6 Measured and calculated $Q$-factors of the cantilever at different geometries. The calculation was done on an Au/Cr covered silicon cantilever after considering air damping, support loss and TED effect.

Besides, the actuation voltage applied to PZT film was found as low as 0.2 volt. In the proposed structure, resonant frequency of the cantilever can be detected by an on-chip integrated piezoresistive gauge, and the piezoelectric PZT film was only used for actuation. Therefore, it is no doubt that aluminum nitride (AlN) or zinc oxide (ZnO) film can be used as a substitution to PZT film for cantilever actuation. AlN or ZnO film offers better compatibility with CMOS from the fabrication viewpoint because it can be deposited by sputter at low temperature or even room temperature [23, 24]. However, actuation voltage of about 2~3 volts is required due to low piezoelectric coefficient $d_{33}$ of the AlN or ZnO film, while it is still acceptable for MEMS-CMOS integration.

### B. Under Reduced Pressure

Under atmospheric pressure, air damping is the predominant mechanism for energy dissipation [17]. Air damping becomes less effective if pressure goes down, and then it is easier to identify the effects of other energy dissipation mechanisms. Solid marks in Fig. 7 shows measured *Q*-factors of a 30 μm-wide 100 μm-long cantilever under various pressures. *Q*-factor was 1113 under the pressure of 101.2 KPa (local atmospheric pressure). It increases markedly when pressure reduces. Under the pressure of 35 Pa, *Q*-factor was measured as 7279. Fig. 8 shows a sharp resonant peak of this cantilever under the pressure of 35 Pa.

*Q*-factor dependence on pressure can be calculated using equation (2)~(5), in which equation (3) can not give satisfied solution to air damping in full pressure range because it based on viscous approximation [12]. Thus, equation (3) should be replaced by equation (6) as shown below when pressure goes down to molecular region.

$$Q_{air\text{-}molecular} = \sqrt{\frac{3\pi}{128} \cdot \frac{R_{gas} \cdot T}{M}} \cdot k_n^2 \cdot \sqrt{\rho \cdot E} \cdot \frac{t^2}{L^2} \cdot \frac{1}{P_{air}} \quad (6)$$

where $R_{gas}$ is the gas constant of the surrounding air, *T* is the absolute temperature, and *M* is the gas molecules. For air, *M* equals 28.964 gr/mol. Solid line and dot line in Fig. 7 shows calculated *Q*-factors under various pressures. Clearly, the experimental *Q*-factor agree well with the theoretical analysis. It confirms our above conclusion under the reduced pressure region that the energy dissipation from the PZT film and the actuator was effectively compressed by separating the actuator form the resonant structure. In addition, both the measured and the calculated *Q*-factor saturated at 35 Pa as shown in Fig. 7, it can be seen approximately as the intrinsic *Q*-factor (*Q*=7279).

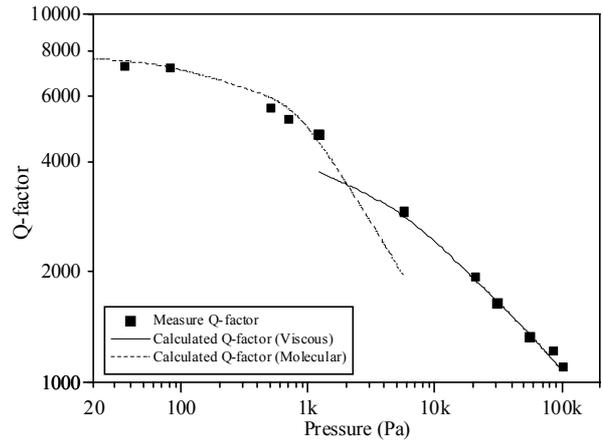

Fig. 7 Measured and calculated *Q*-factors of the cantilever under various pressures (cantilever width: 30 μm, length: 100 μm). The solid line and the dot line was calculated using viscous and molecular approximation respectively in oscillating sphere model for air damping effect.

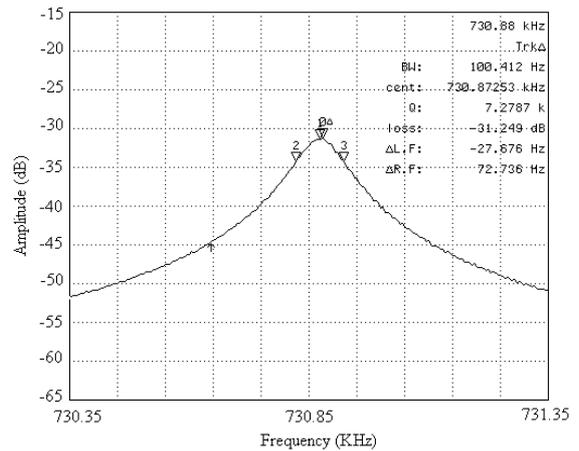

Fig. 8 Resonant peak of the cantilever under the pressure of 35 Pa (cantilever width: 30 μm, cantilever length: 100 μm).

Intrinsic *Q*-factor of a cantilever is supposed to be further improved at longer cantilever to keep it far away from support loss effect. Fig. 9 shows measured *Q*-factor dependence on pressure at various cantilever lengths. *Q*-factor of 100, 150 and 200 μm-long cantilever was measured as 7279, 7978 and 8496 respectively under the pressure of 35 Pa. However, the calculated *Q*-factors were 7495, 19801 and 32411. The difference between measured and calculated *Q*-factor increased with the cantilever length. It indicates that unforeseen energy dissipation exists in long cantilevers (in short cantilever such as 100 μm, *Q*-factor fit well with the calculation as shown in Fig. 7), and this mechanism is comparable to other mechanism only under reduced pressure (in molecular region). Since Au/Cr covered area became larger with the increase of cantilever length, the most possible reason is believed due to Au/Cr deposition. Sekaric et al. studied the effects of thin metal films on energy dissipation in nano-scale silicon nitride mechanical resonators, and they found that the absence of these metal coatings results in a three to four times higher Q-factor [25]. The reason might be inter-friction between each layer, while further investigations are necessary to clarify it.

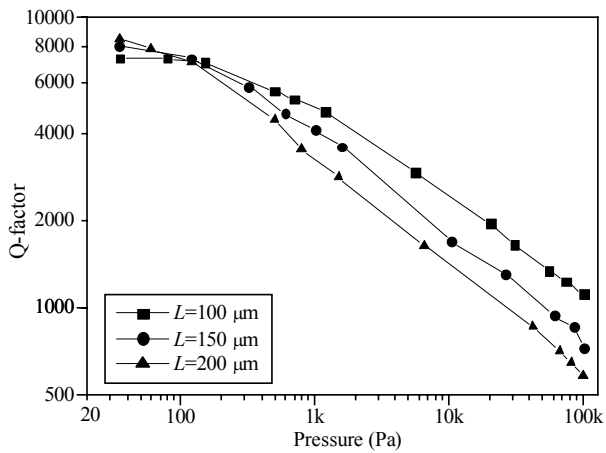

Fig. 9 *Q*-factor dependence on pressure at various cantilever lengths (cantilever width: 30 μm). The measured *Q*-factor under 35 Pa was 7279, 7978 and 8496 for 100, 150 and 200 μm-long cantilever, respectively.

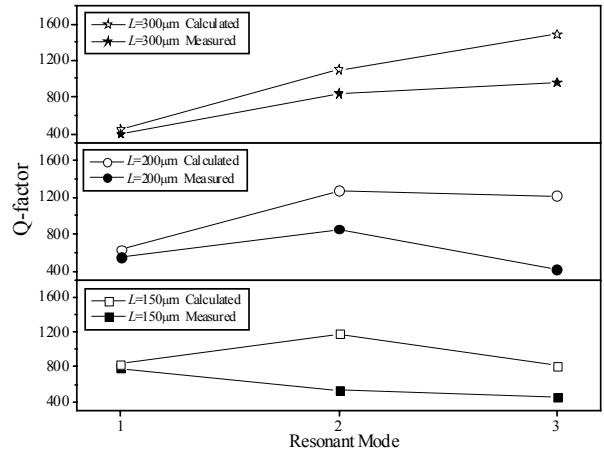

Fig. 10 Measured and calculated Q-factor dependence on resonant mode at different cantilever lengths (cantilever width: 50 μm).

*C. High Mode Vibration*

High-mode vibration can improve mass detect sensitivity of a resonant cantilever under atmospheric pressure by suppressing the air damping effect [12]. Fig. 10 shows measured and calculated *Q*-factor dependence on resonant mode at different cantilever geometries. High mode vibration can be successfully achieved by the proposed structure, and greater *Q*-factor can be obtained as expected for the pursuit of better mass detection sensitivity.

It is noteworthy that in Fig. 10, long cantilever is preferable for high mode vibration. However, the measured *Q*-factors are still lower than the theoretical calculations. High mode vibration results in large vibration amplitude at the position where cantilever and actuator connects. It leads to large vibration amplitude in PZT actuator, which in turn induces additional energy dissipation as analyzed in reference [10]. Besides, large vibration amplitude at the actuation hinge also trends to decrease $Q_{sup}$, because energy may dissipate easily through substrate. For high mode vibration, the actuation position should be close to node-point of the cantilever (the position with no displacement at resonant frequency in theory) to suppress energy dissipation. Fig. 11 shows schematic view of a suggested structure for high mode vibration. This work is still under going. We will report the experimental results in our future publications.

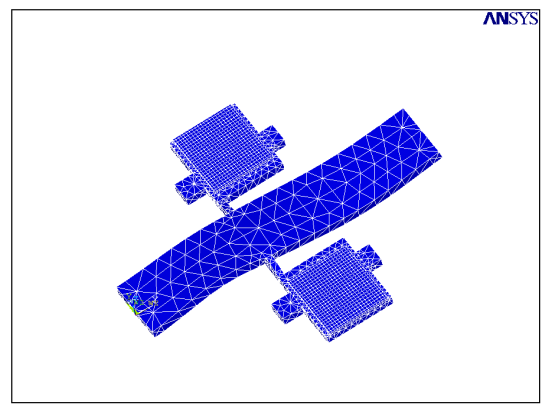

Fig. 11 Schematic view of a suggested structure for high mode vibration (3rd mode as shown in this figure).

## IV. CONCLUSIONS

In this paper, a piezoelectric PZT thin film actuated single crystal silicon microcantilever was proposed for resonant based ultra sensitive mass detection. Excellent *Q*-factor was achieved under both atmospheric pressure and reduced pressures by separating the PZT actuator from the resonant structure to compress the energy dissipation and other negative effects. The proposed structure exhibits attractive integration capabilities with other MEMS components and CMOS circuits for application in parallel sensor system or networked sensor technology.

ACKNOWLEDGMENT

This work was partially sponsored by Japan Society for the Promotion of Science (JSPS) under the Program of Postdoctoral Fellowships for Foreign Researchers (FY-2007).